\pdfoutput=1
\documentclass[12pt,a4paper]{article}
\usepackage{ifthen} % for conditional statements
\newboolean{pdflatex}
\setboolean{pdflatex}{true} % False for eps figures 

\newboolean{articletitles}
\setboolean{articletitles}{true} % False removes titles in references

\newboolean{uprightparticles}
\setboolean{uprightparticles}{false} %True for upright particle symbols

\def\paperauthors{LHCb collaboration} % Leave as is for PAPER, CONF and FIGURE
\def\paperasciititle{Search for the lepton flavour violating decay B+ -> K+ mu- tau+ using Bs2* decays} % Set ASCII title here
\def\papertitle{Search for the lepton flavour violating decay \decay{\Bu}{\Kp\mun\taup} using \bstwost decays} % Latex formatted title
\def\paperkeywords{{High Energy Physics}, {LHCb}} % Comma separated list
\def\papercopyright{\the\year\ CERN for the benefit of the LHCb collaboration} % new since 9/Apr/2018
\def\paperlicence{CC BY 4.0 licence}
\def\paperlicenceurl{https://creativecommons.org/licenses/by/4.0/}

\usepackage[top=1in, bottom=1.25in, left=1in, right=1in]{geometry}

\usepackage{microtype}
\usepackage{lineno}  % for line numbering during review
\usepackage{xspace} % To avoid problems with missing or double spaces after
\usepackage{caption} %these three command get the figure and table captions automatically small

\usepackage{graphicx}  % to include figures (can also use other packages)
\usepackage{color}
\usepackage{colortbl}
\graphicspath{{./figs/},{./fig_final/}} % Make Latex search fig subdir for figures
\DeclareGraphicsExtensions{.pdf,.PDF,png,.PNG}

\usepackage{amsmath} % Adds a large collection of math symbols
\usepackage{amssymb}
\usepackage{amsfonts}
\usepackage{upgreek} % Adds in support for greek letters in roman typeset

\newcommand*\patchAmsMathEnvironmentForLineno[1]{%
\expandafter\let\csname old#1\expandafter\endcsname\csname #1\endcsname
\expandafter\let\csname oldend#1\expandafter\endcsname\csname
end#1\endcsname
 \renewenvironment{#1}%
   {\linenomath\csname old#1\endcsname}%
   {\csname oldend#1\endcsname\endlinenomath}%
}
\newcommand*\patchBothAmsMathEnvironmentsForLineno[1]{%
  \patchAmsMathEnvironmentForLineno{#1}%
  \patchAmsMathEnvironmentForLineno{#1*}%
}
\AtBeginDocument{%
\patchBothAmsMathEnvironmentsForLineno{equation}%
\patchBothAmsMathEnvironmentsForLineno{align}%
\patchBothAmsMathEnvironmentsForLineno{flalign}%
\patchBothAmsMathEnvironmentsForLineno{alignat}%
\patchBothAmsMathEnvironmentsForLineno{gather}%
\patchBothAmsMathEnvironmentsForLineno{multline}%
\patchBothAmsMathEnvironmentsForLineno{eqnarray}%
}

\usepackage{hyperxmp}

\usepackage[pdftex,
            pdfauthor={\paperauthors},
            pdftitle={\paperasciititle},
            pdfkeywords={\paperkeywords},
            pdfcopyright={Copyright (C) \papercopyright},
            pdflicenseurl={\paperlicenceurl}]{hyperref}

\usepackage[colorinlistoftodos,textsize=scriptsize]{todonotes}

\usepackage[all]{hypcap} % Internal hyperlinks to floats.

\usepackage{xspace} 
\usepackage{upgreek}

\def\lhcb   {\mbox{LHCb}\xspace}

\def\MagUp {\mbox{\em Mag\kern -0.05em Up}\xspace}

\ifthenelse{\boolean{uprightparticles}}%
{

 \def\Pmu         {\ensuremath{\upmu}\xspace}                 
 \def\Pnu         {\ensuremath{\upnu}\xspace}                 
                  
 \def\Ppi         {\ensuremath{\uppi}\xspace}

 \def\Ptau        {\ensuremath{\uptau}\xspace}

 \def\Ppsi        {\ensuremath{\uppsi}\xspace}

 \def\PDelta      {\ensuremath{\Delta}\xspace}                 
 \def\PXi         {\ensuremath{\Xi}\xspace}                 
 \def\PLambda     {\ensuremath{\Lambda}\xspace}                 
 \def\PSigma      {\ensuremath{\Sigma}\xspace}                 
 \def\POmega      {\ensuremath{\Omega}\xspace}                 
 \def\PUpsilon    {\ensuremath{\Upsilon}\xspace}

 \def\PB      {\ensuremath{\mathrm{B}}\xspace}                 
                  
 \def\PD      {\ensuremath{\mathrm{D}}\xspace}

 \def\PJ      {\ensuremath{\mathrm{J}}\xspace}                 
 \def\PK      {\ensuremath{\mathrm{K}}\xspace}

 \def\Pb      {\ensuremath{\mathrm{b}}\xspace}                 
 \def\Pc      {\ensuremath{\mathrm{c}}\xspace}

 \def\Pi      {\ensuremath{\mathrm{i}}\xspace}

 \def\Ps      {\ensuremath{\mathrm{s}}\xspace}

 \def\thebaroffset{0.0em}
}
{

 \def\Pmu         {\ensuremath{\mu}\xspace}                 
 \def\Pnu         {\ensuremath{\nu}\xspace}                 
                  
 \def\Ppi         {\ensuremath{\pi}\xspace}

 \def\Ptau        {\ensuremath{\tau}\xspace}

 \def\Ppsi        {\ensuremath{\psi}\xspace}                 
                  
 \mathchardef\PDelta="7101
 \mathchardef\PXi="7104
 \mathchardef\PLambda="7103
 \mathchardef\PSigma="7106
 \mathchardef\POmega="710A
 \mathchardef\PUpsilon="7107
                  
 \def\PB      {\ensuremath{B}\xspace}                 
                  
 \def\PD      {\ensuremath{D}\xspace}

 \def\PJ      {\ensuremath{J}\xspace}                 
 \def\PK      {\ensuremath{K}\xspace}

 \def\Pb      {\ensuremath{b}\xspace}                 
 \def\Pc      {\ensuremath{c}\xspace}

 \def\Pi      {\ensuremath{i}\xspace}

 \def\Ps      {\ensuremath{s}\xspace}

 \def\thebaroffset{0.18em}
}
\newcommand{\offsetoverline}[2][\thebaroffset]{\kern #1\overline{\kern -#1 #2}}%

\makeatletter
\ifcase \@ptsize \relax% 10pt
  \newcommand{\miniscule}{\@setfontsize\miniscule{4}{5}}% \tiny: 5/6
\or% 11pt
  \newcommand{\miniscule}{\@setfontsize\miniscule{5}{6}}% \tiny: 6/7
\or% 12pt
  \newcommand{\miniscule}{\@setfontsize\miniscule{5}{6}}% \tiny: 6/7
\fi
\makeatother

\DeclareRobustCommand{\optbar}[1]{\shortstack{{\miniscule (\rule[.5ex]{1.25em}{.18mm})}
  \\ [-.7ex] $#1$}}

\def\mup        {{\ensuremath{\Pmu^+}}\xspace}
\def\mun        {{\ensuremath{\Pmu^-}}\xspace} % muon negative (\mum is taken)

\def\taup       {{\ensuremath{\Ptau^+}}\xspace}
\def\taum       {{\ensuremath{\Ptau^-}}\xspace}

\def\neu        {{\ensuremath{\Pnu}}\xspace}
\def\neub       {{\ensuremath{\overline{\Pnu}}}\xspace}
\def\neum       {{\ensuremath{\neu_\mu}}\xspace}

\def\neutb      {{\ensuremath{\neub_\tau}}\xspace}
\def\squark    {{\ensuremath{\Ps}}\xspace}

\def\cquark    {{\ensuremath{\Pc}}\xspace}

\def\bquark    {{\ensuremath{\Pb}}\xspace}

\def\pion   {{\ensuremath{\Ppi}}\xspace}

\def\pip    {{\ensuremath{\pion^+}}\xspace}
\def\pim    {{\ensuremath{\pion^-}}\xspace}

\def\kaon    {{\ensuremath{\PK}}\xspace}
\def\KorKbar {\kern \thebaroffset\optbar{\kern -\thebaroffset \PK}{}\xspace}

\def\Kp      {{\ensuremath{\kaon^+}}\xspace}
\def\Km      {{\ensuremath{\kaon^-}}\xspace}

\def\Dbar    {{\ensuremath{\offsetoverline{\PD}}}\xspace}
\def\D       {{\ensuremath{\PD}}\xspace}

\def\DorDbar {\kern \thebaroffset\optbar{\kern -\thebaroffset \PD}\xspace}

\def\Dzb     {{\ensuremath{\Dbar{}^0}}\xspace}
\def\Dp      {{\ensuremath{\D^+}}\xspace}
\def\Dm      {{\ensuremath{\D^-}}\xspace}

\def\DpDm    {\ensuremath{\Dp {\kern -0.16em \Dm}}\xspace}

\def\B       {{\ensuremath{\PB}}\xspace}

\def\BorBbar {\kern \thebaroffset\optbar{\kern -\thebaroffset \PB}\xspace}

\def\Bd      {{\ensuremath{\B^0}}\xspace}

\def\BdorBdbar {\kern \thebaroffset\optbar{\kern -\thebaroffset \Bd}\xspace}
\def\Bu      {{\ensuremath{\B^+}}\xspace}

\def\Bs      {{\ensuremath{\B^0_\squark}}\xspace}

\def\BsorBsbar {\kern \thebaroffset\optbar{\kern -\thebaroffset \Bs}\xspace}

\def\jpsi     {{\ensuremath{{\PJ\mskip -3mu/\mskip -2mu\Ppsi}}}\xspace}

\def\Y#1S{\ensuremath{\PUpsilon{(#1S)}}\xspace}

\def\Lz          {{\ensuremath{\PLambda}}\xspace}

\def\LorLbar     {\kern \thebaroffset\optbar{\kern -\thebaroffset \PLambda}\xspace}

\def\Lb           {{\ensuremath{\Lz^0_\bquark}}\xspace}

\def\BF         {{\ensuremath{\mathcal{B}}}\xspace}

\newcommand{\decay}[2]{\ensuremath{#1\!\to #2}\xspace} 

\def\to                 {\ensuremath{\rightarrow}\xspace}

\def\AT#1     {\ensuremath{A_{\mathrm{T}}^{#1}}\xspace}           % 2

\def\C#1      {\ensuremath{\mathcal{C}_{#1}}\xspace}                       % 9
\def\Cp#1     {\ensuremath{\mathcal{C}_{#1}^{'}}\xspace}                    % 7
\def\Ceff#1   {\ensuremath{\mathcal{C}_{#1}^{\mathrm{(eff)}}}\xspace}        % 9  
\def\Cpeff#1  {\ensuremath{\mathcal{C}_{#1}^{'\mathrm{(eff)}}}\xspace}       % 7
\def\Ope#1    {\ensuremath{\mathcal{O}_{#1}}\xspace}                       % 2
\def\Opep#1   {\ensuremath{\mathcal{O}_{#1}^{'}}\xspace}                    % 7

\newcommand{\nospaceunit}[1]{\ensuremath{\text{#1}}}       
\newcommand{\aunit}[1]{\ensuremath{\text{\,#1}}}       
\newcommand{\tev}{\aunit{Te\kern -0.1em V}\xspace}
\newcommand{\gev}{\aunit{Ge\kern -0.1em V}\xspace}
\newcommand{\mev}{\aunit{Me\kern -0.1em V}\xspace}
\newcommand{\kev}{\aunit{ke\kern -0.1em V}\xspace}
\newcommand{\ev}{\aunit{e\kern -0.1em V}\xspace}
 
\newcommand{\mevc}{\ensuremath{\aunit{Me\kern -0.1em V\!/}c}\xspace}
\newcommand{\gevc}{\ensuremath{\aunit{Ge\kern -0.1em V\!/}c}\xspace}
\newcommand{\mevcc}{\ensuremath{\aunit{Me\kern -0.1em V\!/}c^2}\xspace}
\newcommand{\gevcc}{\ensuremath{\aunit{Ge\kern -0.1em V\!/}c^2}\xspace}
\def\mum  {\ensuremath{\,\upmu\nospaceunit{m}}\xspace}

\def\fb   {\ensuremath{\aunit{fb}}\xspace}
\def\invfb   {\ensuremath{\fb^{-1}}\xspace}

\def\gsim{{~\raise.15em\hbox{$>$}\kern-.85em
          \lower.35em\hbox{$\sim$}~}\xspace}
\def\lsim{{~\raise.15em\hbox{$<$}\kern-.85em
          \lower.35em\hbox{$\sim$}~}\xspace}

\newcommand{\abs}[1]{\ensuremath{\left\|#1\right\|}} % {x}

\def\pt         {\ensuremath{p_{\mathrm{T}}}\xspace}

\def\ptot       {\ensuremath{p}\xspace}

\def\evtgen     {\mbox{\textsc{EvtGen}}\xspace}

\def\geant      {\mbox{\textsc{Geant4}}\xspace}

\def\pythia     {\mbox{\textsc{Pythia}}\xspace}

\def\tell1  {TELL1\xspace}
\def\ukl1   {UKL1\xspace}

\def\OSK {\text{OS}\kaon}
\def\SSK {\text{SS}\kaon}

\def\bstwost{{\ensuremath{B_{s2}^{*0}}}\xspace}

\def\bsone{{\ensuremath{B_{s1}^0}}\xspace}

\def\kmu{\ensuremath{\Kp\mun}\xspace}

\def\kmut{\ensuremath{\Kp\mun t^+}\xspace}

\def\mmsq{\ensuremath{m_{\mathrm{miss}}^2}\xspace}

\def\dmmin{\ensuremath{\Delta m_{\mathrm{min}}}\xspace}

\def\f12{\ensuremath{\frac{1}{2}}\xspace}

\newenvironment{sfhalf}{
  \begin{subfigure}{0.49\textwidth}}
{\end{subfigure}}

\usepackage{subcaption}
\usepackage[separate-uncertainty=true]{siunitx}
\DeclareSIUnit\evolt{e\kern -0.1em V}
\usepackage{physics}
\usepackage{booktabs}
\usepackage[capitalize]{cleveref}
\crefname{section}{Sect.}{Sect.}
\Crefname{section}{Section}{Sections}

\usepackage{multirow}

\usepackage{cite} % Allows for ranges in citations
\usepackage{mciteplus}

\usepackage{longtable} % only for template; not usually to be used in PAPERs

\begin{document}

%%%%%%%%%%%%%%%%%%%%%%%%%
%%%%% Title     %%%%%%%%%
%%%%%%%%%%%%%%%%%%%%%%%%%
\renewcommand{\thefootnote}{\fnsymbol{footnote}}
\setcounter{footnote}{1}

% %%%%%%% CHOOSE TITLE PAGE--------
%\onecolumn
%\input{title-LHCb-INT}
%\input{title-LHCb-ANA}
%\input{title-LHCb-CONF}
%\input{title-LHCb-FIGURE}
% $Id: title-LHCb-PAPER.tex 122889 2018-08-17 17:59:55Z pkoppenb $
% ===============================================================================
% Purpose: LHCb-PAPER journal paper title page template
% Author: 
% Created on: 2010-09-25
% ===============================================================================

%%%%%%%%%%%%%%%%%%%%%%%%%
%%%%%  TITLE PAGE  %%%%%%
%%%%%%%%%%%%%%%%%%%%%%%%%
\begin{titlepage}
\pagenumbering{roman}

% Header ---------------------------------------------------
\vspace*{-1.5cm}
\centerline{\large EUROPEAN ORGANIZATION FOR NUCLEAR RESEARCH (CERN)}
\vspace*{1.5cm}
\noindent
\begin{tabular*}{\linewidth}{lc@{\extracolsep{\fill}}r@{\extracolsep{0pt}}}
\ifthenelse{\boolean{pdflatex}}% Logo format choice
{\vspace*{-1.5cm}\mbox{\!\!\!\includegraphics[width=.14\textwidth]{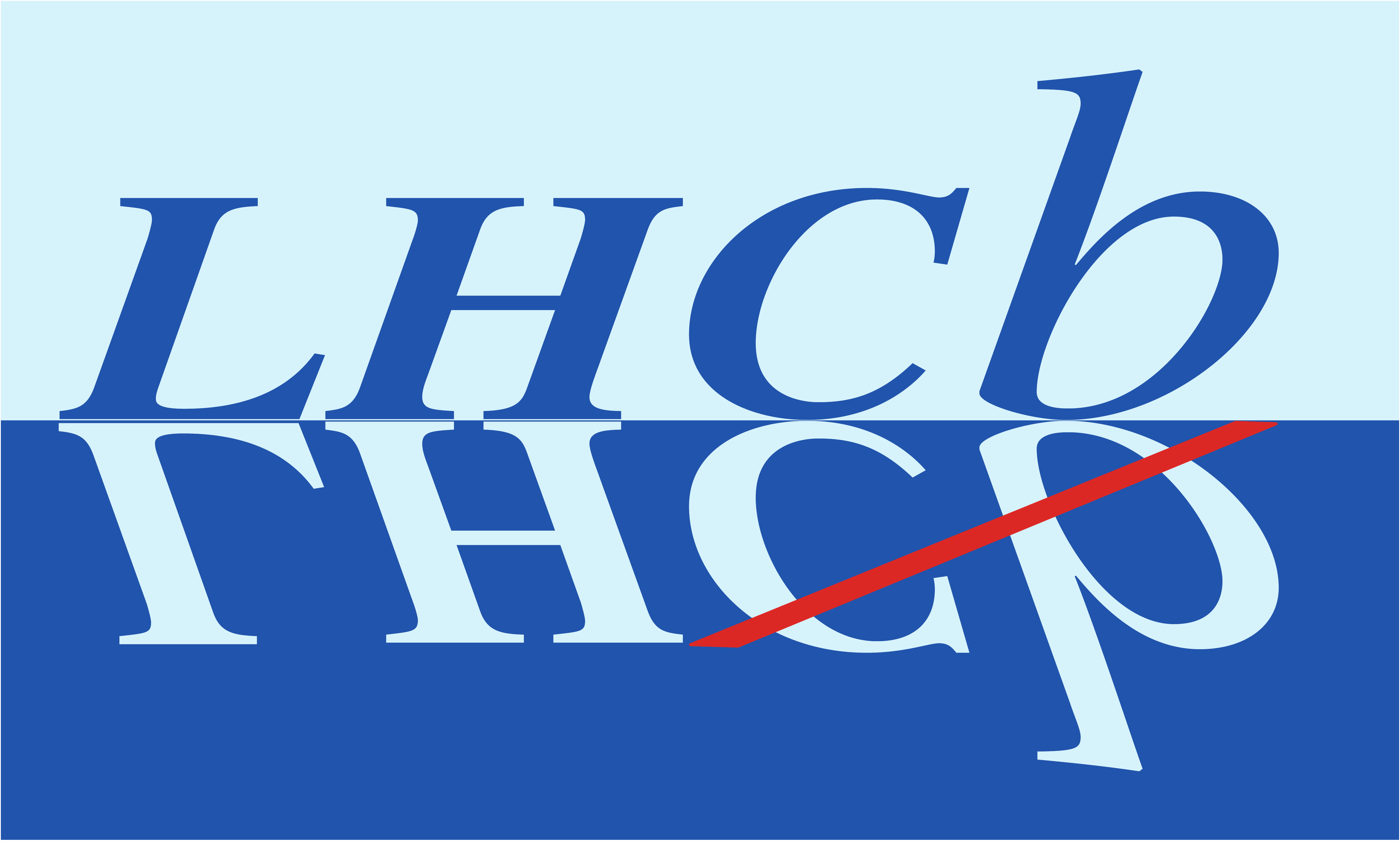}} & &}%
{\vspace*{-1.2cm}\mbox{\!\!\!\includegraphics[width=.12\textwidth]{figs/lhcb-logo.eps}} & &}%
\\
 & & CERN-EP-2020-020 \\  % ID 
 & & LHCb-PAPER-2019-043 \\  % ID 
 & & 9 March 2020 \\ % Date - Can also hardwire e.g.: 23 March 2010
 & & \\
% not in paper \hline
\end{tabular*}

\vspace*{4.0cm}

% Title --------------------------------------------------
{\normalfont\bfseries\boldmath\huge
\begin{center}
% DO NOT EDIT HERE. Instead edit macro in main.tex to keep metadata correct
  \papertitle 
\end{center}
}

\vspace*{2.0cm}

% Authors -------------------------------------------------
\begin{center}
%In the footnote, replace 'paper' by 'Letter' in case of submission to PRL or PLB 
% Edit macro in main.tex to keep metadata correct
\paperauthors\footnote{Authors are listed at the end of this paper.}
\end{center}

\vspace{\fill}

% Abstract -----------------------------------------------
\begin{abstract}
\noindent A search is presented for the lepton flavour violating decay \decay{\Bu}{\Kp\mun\taup} using a sample of proton--proton collisions at centre-of-mass energies of 7, 8, and \SI{13}{\tera\evolt}, collected with the LHCb detector and corresponding to a total integrated luminosity of 9\invfb. The $\tau$ leptons are selected inclusively, primarily via decays with  a single charged particle. The four-momentum of the $\tau$ lepton is determined by using \Bu mesons from $\bstwost\to \Bu \Km$ decays. No significant excess is observed, and an upper limit is set on the branching fraction
  \begin{equation*}
    \BF\qty(\decay{\Bu}{\Kp\mun\taup}) < \num{3.9e-5}\text{ at 90\% confidence level}.
  \end{equation*}
The obtained limit is comparable to the world-best limit.
\end{abstract}

\vspace*{2.0cm}

\begin{center}
  Submitted to
  JHEP
\end{center}

\vspace{\fill}

{\footnotesize 
% Edit macro in main.tex to keep metadata correct
\centerline{\copyright~\papercopyright. \href{\paperlicenceurl}{\paperlicence}.}}
\vspace*{2mm}

\end{titlepage}

%%%%%%%%%%%%%%%%%%%%%%%%%%%%%%%%
%%%%%  EOD OF TITLE PAGE  %%%%%%
%%%%%%%%%%%%%%%%%%%%%%%%%%%%%%%%

%  empty page follows the title page ----
\newpage
\setcounter{page}{2}
\mbox{~}
%\newpage
%
%% Author List ----------------------------
%%  You need to get a new author list!
%\input{LHCb_authorlist.tex}
%
%The author list for journal publications is provided by the Membership Committee shortly after 'approval to go to paper' has been given.
%%It will be made available on the page
%%\verb!http://www.physik.uzh.ch/~strauman/forMemCo/LHCb-PAPER-XXXX-XXX/! .
%It will be sent to you by email shortly after a paper number has beens assigned.
%The author list should be included already at first circulation, 
%to allow new members of the collaboration to verify whether they have been included correctly.
%Occasionally a misspelled name is corrected or associated institutions become full members.
%In that case, a new author list will be sent to you.
%In case line numbering doesn't work well after including the authorlist, try moving the \verb!\bigskip! after the last author to a separate line.
%
%
%The authorship for Conference Reports should be ``The LHCb
%  collaboration'', with a footnote giving the name(s) of the contact
%  author(s), but without the full list of collaboration names.

\cleardoublepage

%\twocolumn
% %%%%%%%%%%%%% ---------

\renewcommand{\thefootnote}{\arabic{footnote}}
\setcounter{footnote}{0}

%%%%%%%%%%%%%%%%%%%%%%%%%%%%%%%%
%%%%%  Table of Content   %%%%%%
%%%%%%%%%%%%%%%%%%%%%%%%%%%%%%%%
%%%% Uncomment next 2 lines if desired
%\tableofcontents
%\cleardoublepage

%%%%%%%%%%%%%%%%%%%%%%%%%
%%%%% Main text %%%%%%%%%
%%%%%%%%%%%%%%%%%%%%%%%%%

\pagestyle{plain} % restore page numbers for the main text
\setcounter{page}{1}
\pagenumbering{arabic}

%% Uncomment during review phase. 
%% Comment before a final submission.
%% \linenumbers

\section{Introduction}

A number of experimental hints of lepton flavour universality violation in the semileptonic transitions \decay{b}{s\ell^+\ell^-}~\cite{LHCb-PAPER-2017-013,LHCb-PAPER-2019-009,LHCb-PAPER-2019-040} and \decay{b}{c\ell^-\neub_{\ell}}~\cite{Lees:2013uzd,Huschle:2015rga,LHCb-PAPER-2017-035,LHCb-PAPER-2017-027,LHCb-PAPER-2017-017,LHCb-PAPER-2015-025} have recently been found.\footnote{The inclusion of charge-conjugate processes is implied throughout.} In general, physics beyond the Standard Model that generates lepton flavour non-universality is likely to also produce direct lepton flavour violation~\cite{Glashow:2014iga}. Theoretical models seeking to simultaneously explain all these anomalies, for example with a vector leptoquark, often lead to relatively large branching fractions for the decays \mbox{\decay{\B}{\kaon \mu^{\pm} \tau^{\mp}}~\cite{Barbieri:2015yvd,Bordone:2017bld,Bordone:2018nbg,Duraisamy:2016gsd,DiLuzio:2017vat,DiLuzio:2018zxy}}.

The branching fractions for the two $\mu\tau$ charge combinations are not in general the same, as they depend on the details of the physics mechanism producing the decay. In this paper, we present a search for the decay \decay{\Bu}{\Kp\mun\taup}. From an experimental point of view, this combination is preferred over \decay{\Bu}{\Kp\mup\taum} as it has a lower background from semileptonic \decay{\B}{\Dbar X \mup \neum} decays, because Cabibbo-favoured decays of the charm meson are likely to lead to kaons of the same charge as the muon. An upper limit on the branching fraction for the signal decay has been previously set by the BaBar collaboration~\cite{Lees:2012zz} \mbox{$\BF\qty(\decay{\Bu}{\Kp\mun\taup}) < \num{2.8e-5}$} at 90\% confidence level (CL).

We reconstruct the full four-momentum of the $\tau$ lepton using \Bu mesons from the decay \decay{\bstwost}{\Bu\Km}, which amounts to about 1\% of \Bu production. By reconstructing the decay vertex of the \Bu meson from the $\Kp\mun$ pair and the momentum of the \Km meson, it is possible to determine the momentum of the \Bu meson up to a quadratic ambiguity by imposing mass constraints on the \bstwost and \Bu mesons~\cite{Stone:2014mza}. This technique was first used to study relative branching fractions in \decay{\Bu}{\Dzb X \mup\nu} decays~\cite{LHCb-PAPER-2018-024}. We then search for a peak in the missing-mass squared distribution corresponding to the $\tau$ mass squared, $m_{\tau}^2$.  Even signal \Bu mesons not coming from a \bstwost decay show a peak at $m_{\tau}^2$. We  account for the contribution of these non-\bstwost candidates in the analysis. The $\tau$ leptons are selected inclusively, as we only require one additional charged track near the $\Kp\mun$ pair to help discriminate against background. To normalise the branching fraction, we use the decay \decay{\Bu}{\jpsi \Kp}, with \decay{\jpsi}{\mup\mun}. The normalisation channel is also used to quantify the contributions from \bstwost decays, as well as non-\bstwost candidates with nearby kaons.

In addition to providing the missing-mass discriminating variable, this method allows us to study the control sample composed of same-sign $\Bu\Kp$ decays, which does not include any \bstwost component. We use this sample to optimise the signal selection, and motivate our description of the background missing-mass shape.

\section{Detector, data samples, and simulation}

The \lhcb detector~\cite{LHCb-DP-2008-001,LHCb-DP-2014-002} is a single-arm forward
spectrometer covering the \mbox{pseudorapidity} range $2<\eta <5$,
designed for the study of particles containing \bquark or \cquark
quarks. The detector includes a high-precision tracking system
consisting of a silicon-strip vertex detector surrounding the $pp$
interaction region, a large-area silicon-strip detector located
upstream of a dipole magnet with a bending power of about
$4{\mathrm{\,Tm}}$, and three stations of silicon-strip detectors and straw
drift tubes placed downstream of the magnet.
The tracking system provides a measurement of the momentum, \ptot, of charged particles with
a relative uncertainty that varies from 0.5\% at low momentum to 1.0\% at 200\gev.\footnote{Natural units with $c = 1$ are used throughout.}
The minimum distance of a track to a primary $pp$ interaction  vertex (PV), the impact parameter, 
is measured with a resolution of $(15+29/\pt)\mum$,
where \pt is the component of the momentum transverse to the beam, in\,\gev.
Different types of charged hadrons are distinguished using information
from two ring-imaging Cherenkov detectors. 
Muons are identified by a
system composed of alternating layers of iron and multiwire
proportional chambers.
The online event selection is performed by a trigger, 
which consists of a hardware stage, based on information from the calorimeter and muon
systems, followed by a software stage, which applies a full event
reconstruction.
At the hardware trigger stage, events are required to have a muon with high \pt or a
  hadron, photon or electron with high transverse energy deposited in the calorimeters. 
  The software trigger requires a two-, three- or four-track
  secondary vertex with a significant displacement from any primary vertex.

We use data samples collected from 2011 to 2018, at centre-of-mass energies of 7, 8, and \SI{13}{\tera\evolt}, corresponding to an integrated luminosity of 9\invfb. We model signal and normalisation decays using simulation. In the simulation, $pp$ collisions are generated using
\pythia~\cite{Sjostrand:2006za,*Sjostrand:2007gs} with a specific \lhcb
configuration~\cite{LHCb-PROC-2010-056}.  Decays of hadronic particles
are described by \evtgen~\cite{Lange:2001uf}. The
interaction of the generated particles with the detector, and its response,
are implemented using the \geant
toolkit~\cite{Allison:2006ve, *Agostinelli:2002hh} as described in
Ref.~\cite{LHCb-PROC-2011-006}.

For the signal, we consider both a phase space model and variations of the decay kinematics with effective operators for the \decay{b}{s\mup\taum} interaction and their corresponding Wilson coefficients using the distributions from Ref.~\cite{Becirevic:2016zri} and the form factors from Ref.~\cite{Ball:2004ye}. The branching fraction limit is determined for various hypotheses: for the phase-space decay, for a decay via the vector or axial-vector operators $\mathcal{O}_9^{(')}$ or $\mathcal{O}_{10}^{(')}$, and for a decay using the scalar or pseudoscalar operators $\mathcal{O}^{(')}_S$ or $\mathcal{O}^{(')}_P$~\cite{Becirevic:2016zri}.

\section{Selection and missing mass calculation}

The selection of \Bu candidates begins with a \kmu pair with an invariant mass \mbox{$m_{\kmu} > \SI{1800}{\mega\evolt}$} to reduce background from semileptonic charm decays. The \Kp and \mun candidates are formed from high-quality tracks consistent with kaon and muon hypotheses and inconsistent with being produced at any PV in the event. The \kmu vertex must be of high quality and well separated from any PV.

To better separate signal candidates with $\tau$ leptons from background, we require an additional track, labelled $t^+$, with charge opposite to that of the muon. By adding this third track, we also fully reconstruct the normalisation mode \decay{\Bu}{\jpsi\Kp}, with \decay{\jpsi}{\mup\mun}. Many background candidates are expected to come from \B-meson decays of the form \mbox{\decay{\B}{\Dbar\qty( \decay{}{\Kp X \mun} ) \Kp Y}}, where $X$ and $Y$ refer to any number of additional particles. In these cases the kaon originating from the \Dbar meson is assigned as the additional track. Since only approximately 2\% of $\tau$ decays contain a charged kaon, we apply particle identification requirements so that the track is unlikely to be a charged kaon. Events in which a candidate \decay{\taup}{\pip\pim\pip\neutb} decay is found are not used in this search to avoid overlap with ongoing searches at LHCb exclusively using this decay channel. In addition, events in which we find multiple candidates are not used in this analysis. These requirements do remove signal with multi-prong $\tau$ decays, with an overall loss of less than 3\%. Multiple candidate events are more likely to come from background, however.
 We split the data samples into signal and normalisation regions based on the invariant mass of the $\kmu t^+$ triple, using the muon hypothesis for the third track. Candidates with $m_{K\mu\mu} < \SI{4800}{\mega\evolt}$ fall into the signal region, while candidates with $5180 < m_{K\mu\mu} < \SI{5380}{\mega\evolt}$ and $\abs{m_{\mu\mu} - m_{\jpsi}} < \SI{40}{\mega\evolt}$ fall into the normalisation region.

The \Bu candidate direction is estimated using the PV and \kmu vertex positions. We next consider prompt tracks, \emph{i.e.} those that are consistent with being produced at that PV. Those tracks identified as kaons, with a charge opposite to that of the kaon in the \kmu pair and a small perpendicular momentum relative to the \Bu candidate direction, are combined with the \Bu candidates to form \bstwost candidates. We refer to this sample as the opposite-sign kaon (\OSK) sample. Additionally, we select a control sample, referred to as same-sign kaon (\SSK) sample, by adding prompt kaons of the same sign as the kaon in the \kmu pair.

From Ref.~\cite{LHCb-PAPER-2018-024}, the two \B-meson energy solutions are
\begin{align}
  E_B &= \frac{\Delta^2}{2E_K}\frac{1}{1 - \qty(p_K/E_K)^2 \cos^2\theta} \qty[ 1 \pm \sqrt{d} ], \text{ where}\\
  d &= \frac{p_K^2}{E_K^2} \cos^2\theta - \frac{4m_B^2p_K^2\cos^2\theta}{\Delta^4}\qty(1 - \frac{p_K^2}{E_K^2} \cos^2\theta ), \\
\Delta^2 &= m_{BK}^2 - m_{B}^2 - m_K^2,
\end{align}
where $m_{BK} = m_{\bstwost}$ is the assumed $\Bu\Km$ mass, $p_K$ and $E_K$ are the reconstructed prompt kaon momentum and energy, and $\theta$ is the laboratory frame angle between the prompt kaon and $B$-meson directions. The missing four-momentum of the $\tau$ lepton, $P_{\text{miss}}$, is then reconstructed as $P_B - P_{\kmu}$, where $P_B$ and $P_{\kmu}$ are the four-momenta of the \B meson and $\kmu$ pair. The missing mass squared is calculated using the lowest energy, real solution for which the resulting missing energy is greater than the reconstructed energy of the third track under a pion mass hypothesis. With this choice, we correctly reconstruct the energy of signal decays in simulation in more than 75\% of cases. About 9\% of all signal decays have no such solution and are lost.  Both signal and normalisation candidates, as well as the \SSK control-sample candidates, are required to pass this procedure. Candidates in the signal region are additionally required to have the residual missing mass squared, defined as the four-momentum difference of the \B meson and $\kmu t^+$ triple, $\qty(P_B - P_{\kmu} - P_{t})^2$, greater than \SI{-0.5}{\giga\evolt\squared}. This requirement removes background and only poorly reconstructed signal candidates which do not peak at the $\tau$ mass squared. The minimum mass difference, defined in Ref.~\cite{LHCb-PAPER-2018-024} as
\begin{equation}
  \dmmin = \sqrt{ m_B^2 + m_K^2 + 2m_B\sqrt{ p_K^2\sin^2\theta + m_K^2} } - m_B - m_K,
\end{equation} 
is required to be greater than \SI{30}{\mega\evolt}. This removes contributions from \bsone and \mbox{\decay{\bstwost}{\B^{*+} \Km}} decays, as well as background in which a kaon from the \B decay is wrongly associated to the primary vertex.

\begin{figure}[htb]
\begin{sfhalf}
\includegraphics[width=\textwidth]{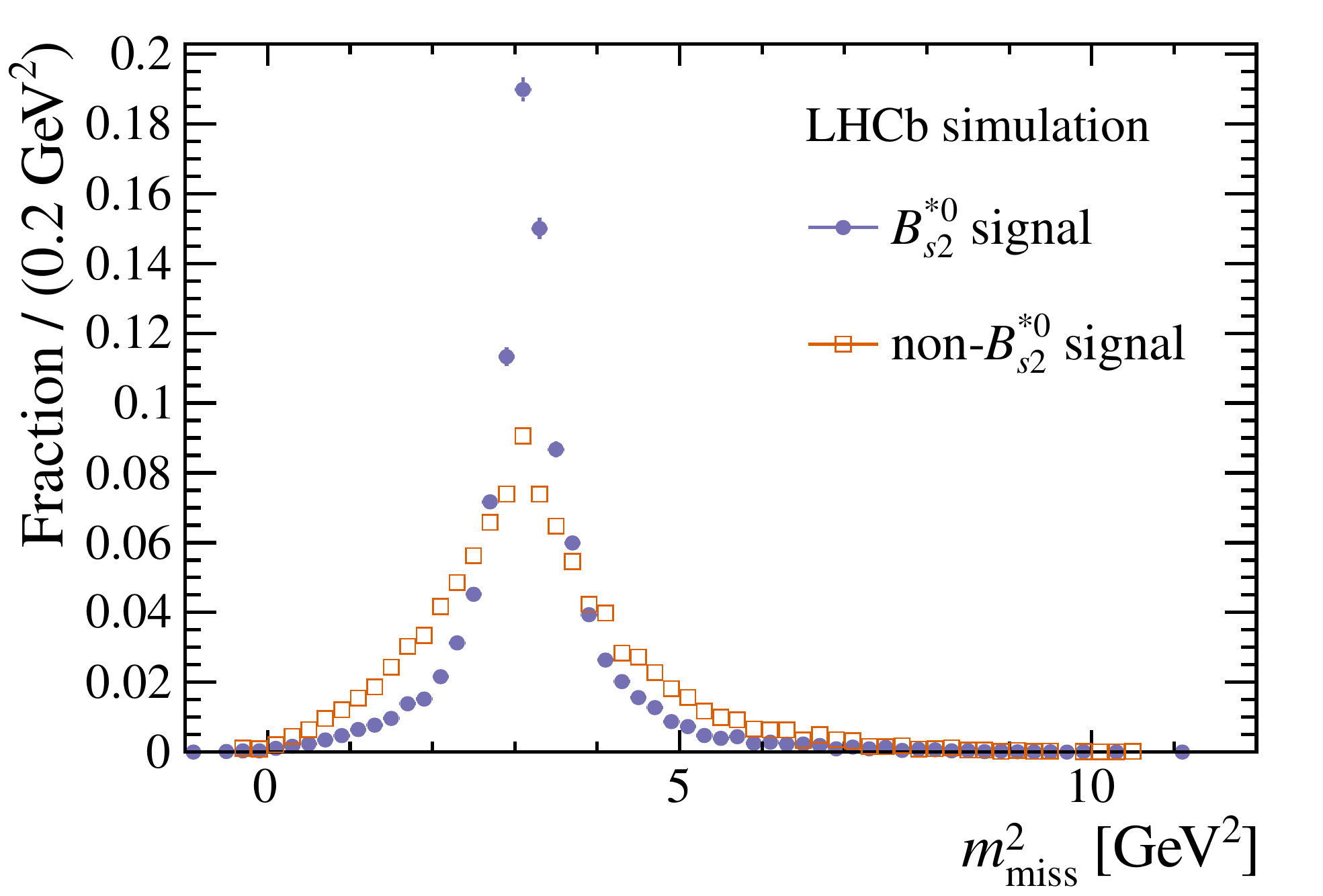}
\end{sfhalf}
\begin{sfhalf}
\includegraphics[width=\textwidth]{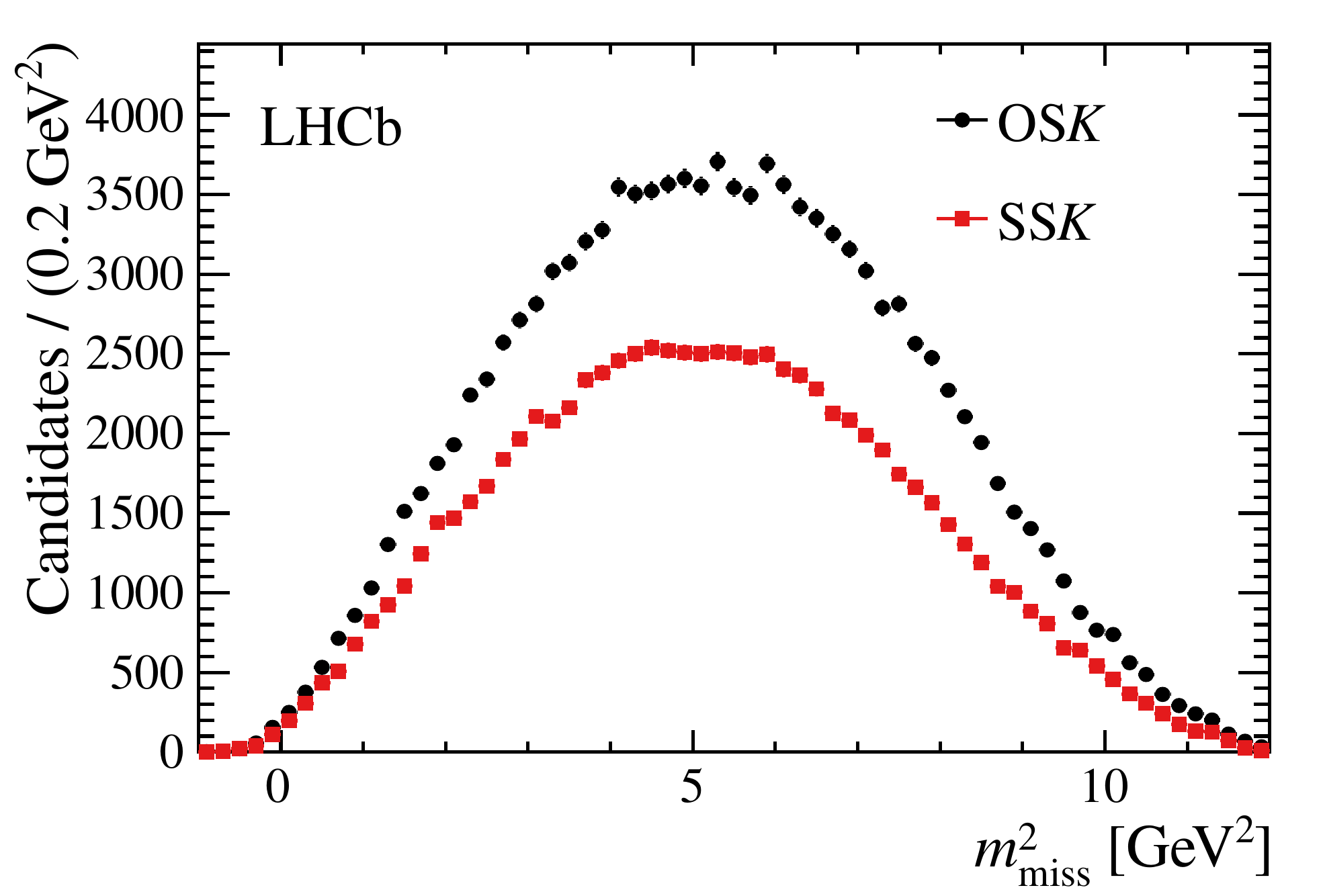}
\end{sfhalf}
\caption{Missing mass squared, \mmsq, distributions for (left) simulated signal \decay{\Bu}{\Kp\mun\taup} decays and (right) all selected candidates in data before applying the signal optimisation described in \cref{sec:signal}.\label{fig:mmsq_raw}}
\end{figure}

Missing-mass distributions for the signal simulation and the full data sample after the above selection are shown in \cref{fig:mmsq_raw}. All signal decays, whether they come from a \bstwost meson or not, peak at the known $m_\tau^2$, however the non-\bstwost candidates have a much wider peak than the \bstwost ones. The data distributions are shown for both the \OSK and \SSK samples. They have similar shapes with a broad hump centred near \SI{5}{\giga\evolt\squared}. We note that the \OSK sample has a higher yield than the \SSK; this excess has been observed in both fully and partially reconstructed decays~\cite{LHCb-PAPER-2012-030,LHCb-PAPER-2018-024}.

\section{Normalisation}

We determine the yield of the normalisation decay, as well as the relative efficiency of the signal modes with respect to the normalisation mode, separately for each data-taking year. For the normalisation mode, we determine the inclusive yield of \decay{\Bu}{\jpsi\Kp} decays, whether or not they originate from a \bstwost meson, by a binned maximum-likelihood fit to the \kmut mass distribution, where we assign the muon mass hypothesis to the third track. The signal is described with a Gaussian distribution, and the background with a linear model. 

\begin{figure}[htb]
\begin{sfhalf}
  \includegraphics[width=\textwidth]{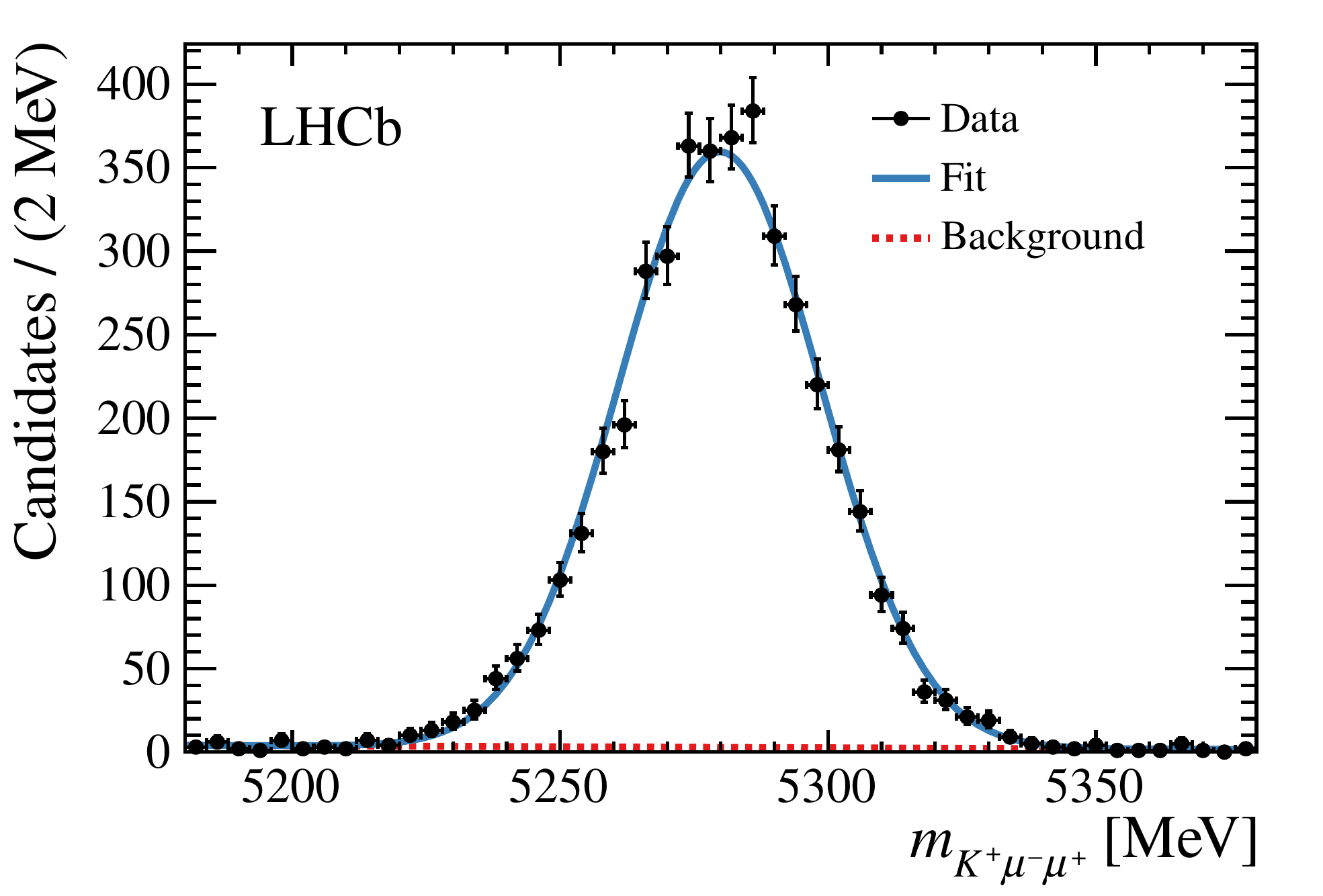}
\end{sfhalf}
\begin{sfhalf}
  \includegraphics[width=\textwidth]{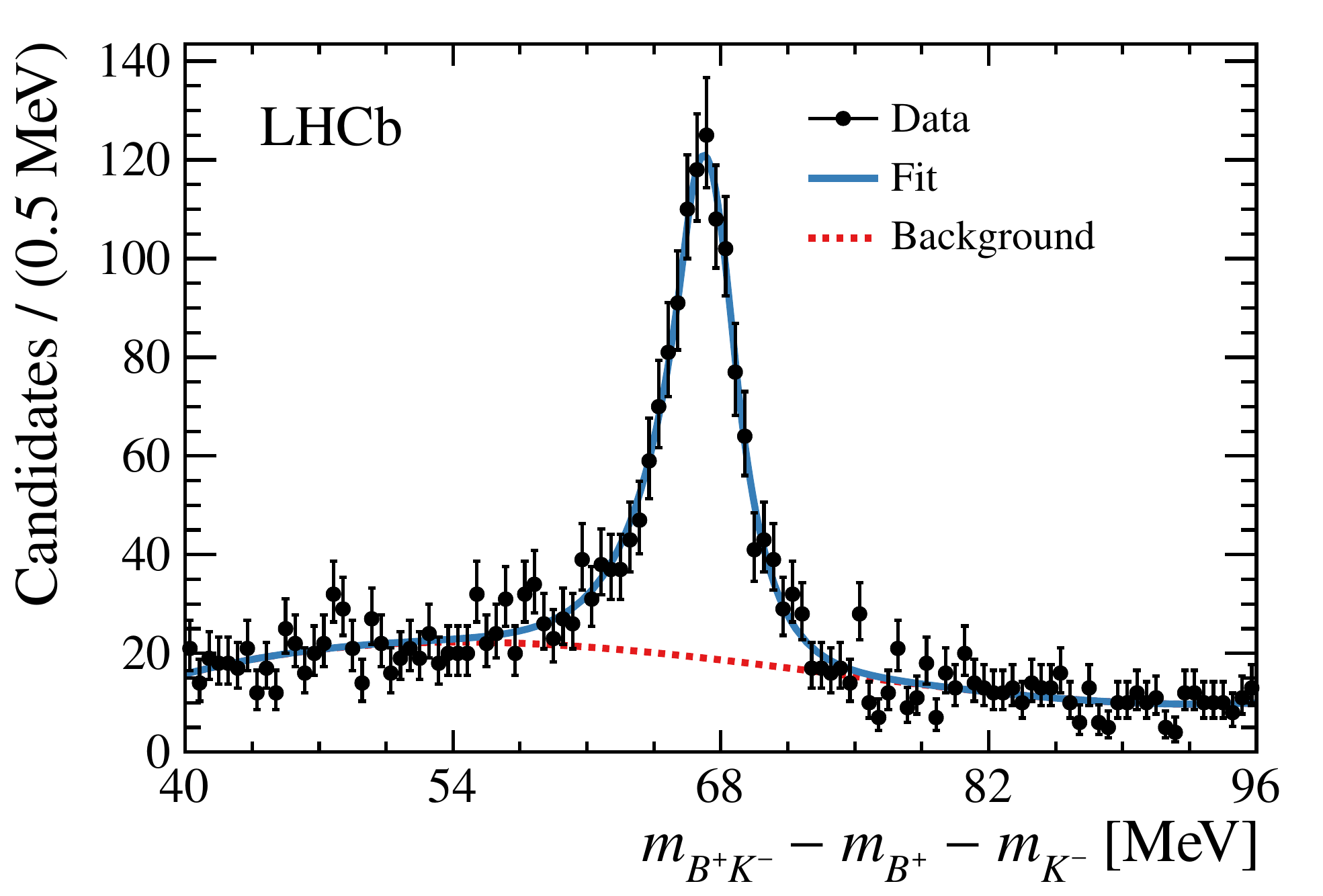}
\end{sfhalf}
\caption{Distributions of normalization candidates in (left) mass, $m_{\Kp\mun\mup}$, and (right) the mass difference, $m_{\Bu\Km} - m_{\Bu} - m_{\Km}$. The result of each fit is shown as a solid line, with the background component as a dashed line.\label{fig:fit_jpsi}}
\end{figure}

We determine the fraction of the normalisation candidates coming from \bstwost decays using a \kmut mass fit for the combined-years data sample using the same model as the separated-years samples, along with a binned maximum-likelihood fit to the measured mass-difference distribution $m_{\Bu\Km} - m_{\Bu} - m_{\Km}$ around the \bstwost peak. For the latter fit, we describe the signal peak with a Gaussian core that transitions to an exponential tail on each side, and we model the background with a third-degree polynomial. The results of these fits are shown in \cref{fig:fit_jpsi}. The total data sample contains \num{4240 \pm 70} \decay{\Bu}{\jpsi\Kp} decays; the fraction originating from \bstwost decays is $f_\bstwost = \SI{25.4 \pm 1.8}{\percent}$, where the uncertainty combines the statistical and systematic uncertainties from the choice of fit function. The year-to-year variation is not found to be statistically significant, so we use the value obtained from the combined dataset for all years.

The relative efficiency of the signal and normalisation modes is determined using simulation with corrections from data. For \bstwost decays the relative efficiencies in different years average around 30\%, with an absolute year-to-year variation of less than 3\%. Different signal decay models change the relative efficiency by approximately 10\%, with the decays via scalar and pseudoscalar operators having a lower overall efficiency. Signal events in which the \Bu meson does not originate from a \bstwost decay have a lower selection efficiency, primarily because fewer of these candidates pass the residual missing-mass requirement and fall into the missing-mass fit range. Using simulation, we derive an additional efficiency factor for this signal component of $r_{\text{non-\bstwost}} = 0.849 \pm 0.007$.

\section{Multivariate signal selection}
\label{sec:signal}

We further improve the signal selection using a Boosted Decision Tree (BDT) classification with the Adaboost algorithm~\cite{AdaBoost}. The BDT inputs are primarily chosen to distinguish additional tracks coming from signal $\tau$ lepton decays from various sources of background. Some examples are semileptonic $b$-hadron decays to charm where the charm hadron produces a kaon with charge opposite that of the muon, or $b$-hadron decays where the muon is produced in the semileptonic decay of a child charm hadron. The background training sample is taken from the \SSK sample in the \mmsq region around $m_\tau^2$. This focuses the training on the sources of background which fall near the signal peak. We describe the signal with simulation samples that include only \bstwost decays; the effect of the BDT on non-\bstwost signal simulation is then estimated separately. The training makes use of different topological reconstructions of the \kmut triple: in addition to the signal selection, we also first combine either the kaon and the track or the muon and the track into a pair before adding the third particle. The pair masses and the flight distance of the pair in each topology help to distinguish the signal from background, for instance when the pair comes from a charm hadron decay. We also include the flight distance of the $\tau$, which we reconstruct as the distance along the $\tau$ trajectory found in the missing-mass calculation from the \kmu vertex to the point of closest approach of the third track. 

The result of a separate isolation discriminant is included to reduce background with additional charged tracks; this discriminant is trained to distinguish additional tracks belonging to the same $b$-hadron decay from other tracks in the event based on kinematic and topological variables.
We perform the rest of the analysis in four bins of the signal optimisation BDT output, keeping about 70\% of all simulated \bstwost signal candidates and about 40\% of non-\bstwost signal candidates. The bins are chosen by optimising the expected upper limit using a number of background events derived from the \OSK and \SSK \mmsq sidebands.

\section{Background studies}
\label{sec:background}

The background in this analysis is composed of a large number of different partially reconstructed $b$-hadron decays. None of them, however, produce a narrow peak in \mmsq. Only \Bu mesons produced from \bstwost decays have a resolution comparable to the signal. Furthermore, if there is more than one missing particle then the true missing-mass distribution will be much wider than the expected signal peak. Charm hadrons have masses close to the $\tau$ mass, however there is no Standard Model decay \mbox{\decay{\Bu}{ \Kp\mun \Dp }}. Because of their low branching fraction, we are not sensitive to decays such as \mbox{\decay{\Bu}{\Kp\pim \Dp}}, where the pion is misidentified as a muon. We expect that the missing-mass distribution, summed over many different background components, is smooth, and we model it as a polynomial.

These assumptions are tested using simulation and data. We produce fast simulation samples with RapidSim~\cite{Cowan:2016tnm} of a number of potential exclusive background sources from \Bu, \Bd, \Bs, and \Lb hadrons; the true missing-mass distributions for these decays are smeared to estimate their shapes in data. No sign of any sharply peaking component is found. In data we consider a number of different control samples, namely all possible $K\mu t$ charge combinations in both \OSK and \SSK samples, excluding the signal selection of $\Kp\mun t^+$ in the \OSK sample. There is no sign of any narrow peak in any of the distributions, even after applying a tight requirement on the BDT output.

Maximum-likelihood fits to the \SSK sample using polynomials of different degrees in the restricted \mmsq range from \SIrange{1}{6}{\giga\evolt\squared} are used to study the background shape in more detail. The optimal number of free polynomial parameters in the most signal-like BDT output bin, based on the best-fit value of $-2\log\mathcal{L}$, penalised by one for each additional parameter, is four. 
We further study the effect of background modelling by performing a large number of pseudoexperiments, both background-only and with injected signal at branching fractions of \num{1e-5} and \num{2e-5}. In these studies, we first fit a background model of some polynomial degree to one of the control samples. From this background model we generate many pseudodatasets that we fit with a model of a different degree. Based on these studies, we take into account the systematic uncertainty due to the background modelling by reporting the weakest limit using background descriptions of third, fourth, or fifth degree polynomials, all of which well describe the background shapes in the pseudoexperiments.

\section{Fit description}
\label{sec:fit}

We search for the $\Kp\mun\taup$ missing-mass peak with an unbinned maximum-ikelihood fit simultaneously in four bins of BDT output in the \OSK $\Kp\mun t^+$ signal channel. The fit is performed in the missing-mass range $1 < \mmsq < \SI{6}{\giga\evolt\squared}$. The parameter of interest is the branching fraction $\BF\qty(\decay{\Bu}{\Kp\mun\taup})$. We describe the \mmsq shape for the signal component with a generalized hyperbolic distribution with shape parameters obtained from simulation. Two signal shapes are used: one for \bstwost decays, and one for the wider non-\bstwost contribution. We determine the shapes separately in each bin of BDT response.  The signal decay model does not significantly affect the signal missing-mass shape. The background is described by polynomial functions which vary independently in each BDT output bin.

We base the normalization of the signal components on the yields of the \decay{\Bu}{\jpsi\Kp} decays determined in data year-by-year.  We combine this together with the relative efficiencies, $\varepsilon_{\text{rel}}$; the known \decay{\Bu}{\jpsi\Kp} with \decay{\jpsi}{\mup\mun} combined branching fraction, abbreviated as $\BF\qty(\jpsi\Kp)$; and the parameter of interest to derive a total number of \decay{\Bu}{\Kp\mun\taup} signal decays. This total is divided between \bstwost and non-\bstwost decays based on the observed fraction in the normalization channel, and then distributed across the four BDT bins. This gives yields in each BDT bin $j$ of 
\begin{align}
  N_j\qty( \decay{\Bu}{\Kp\mun\taup} | \bstwost ) = {}& \varepsilon_{\bstwost,j} \frac{ \BF\qty(\Kp\mun\taup) }{ \BF\qty(\jpsi\Kp) }f_{\bstwost} \times {} \nonumber \\
  &\sum_{i\in \text{years}} \varepsilon_{\text{rel},i} N_i\qty(\jpsi\Kp), \label{eqn:bstwostnorm} \\
  N_j\qty( \decay{\Bu}{\Kp\mun\taup} | \text{non-\bstwost} ) = {}& \varepsilon_{\text{non-\bstwost},j}\frac{ \BF\qty(\Kp\mun\taup) }{ \BF\qty(\jpsi\Kp) } \qty(1 - f_{\bstwost}) \times {} \nonumber \\
  &\sum_{i \in\text{years}} \varepsilon_{\text{rel},i} r_{\text{non-\bstwost}} N_i\qty(\jpsi\Kp),\label{eqn:nobstwostnorm} 
\end{align}
where $\varepsilon_{\bstwost,j}$ and $\varepsilon_{\text{non-\bstwost},j}$ are the separate efficiencies for each signal component to be found in BDT bin $j$.
The main parameters of the fit are thus the \decay{\Bu}{\Kp\mun\taup} branching fraction, four parameters for the background normalisation in each BDT bin, and up to five parameters describing the polynomial background shapes in each BDT bin.

The largest systematic uncertainty comes from the choice of background model. The fifth degree background description obtains the weakest limit among the tested background models. We include the effects of other systematic uncertainties using Gaussian-constrained nuisance parameters. These nuisance parameters modify the normalisation yield, the relative efficiency of the signal and normalisation channels, the signal yield in each BDT bin, and the signal shapes. The largest effects come from the modelling of the kinematics of \bstwost decays in simulation, which results in 5\% changes in the relative efficiency and in the signal fractions in each bin of BDT response. The relative statistical uncertainty of the \bstwost fraction taken from the normalisation channel is also approximately 5\%. Altogether, the total effect of these systematic uncertainties on the final limit is small, at the $10^{-6}$ level.

\section{Results and conclusion}

\begin{figure}[tbp]
  \begin{sfhalf}
    \includegraphics[width=\textwidth]{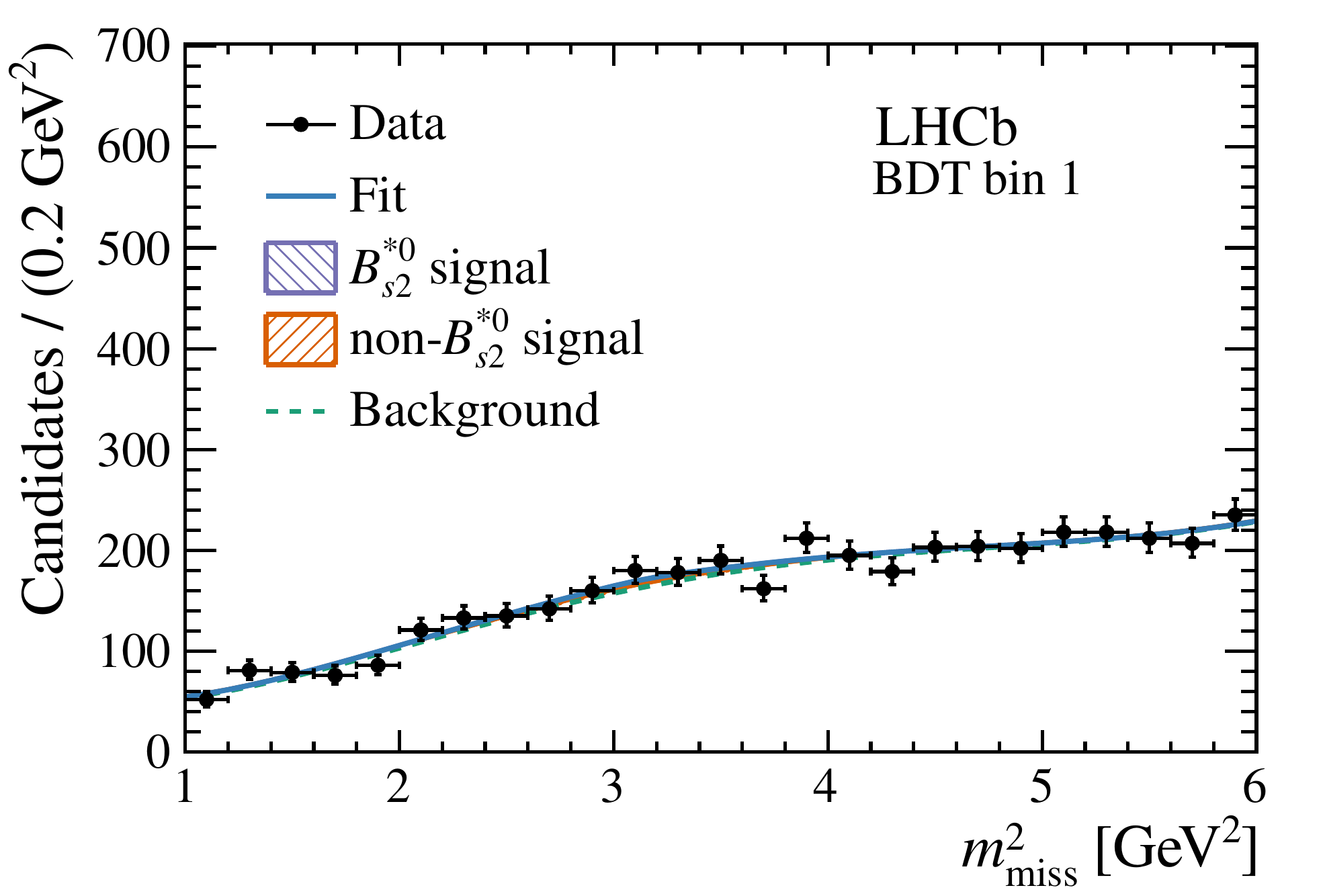}
  \end{sfhalf}
  \begin{sfhalf}
    \includegraphics[width=\textwidth]{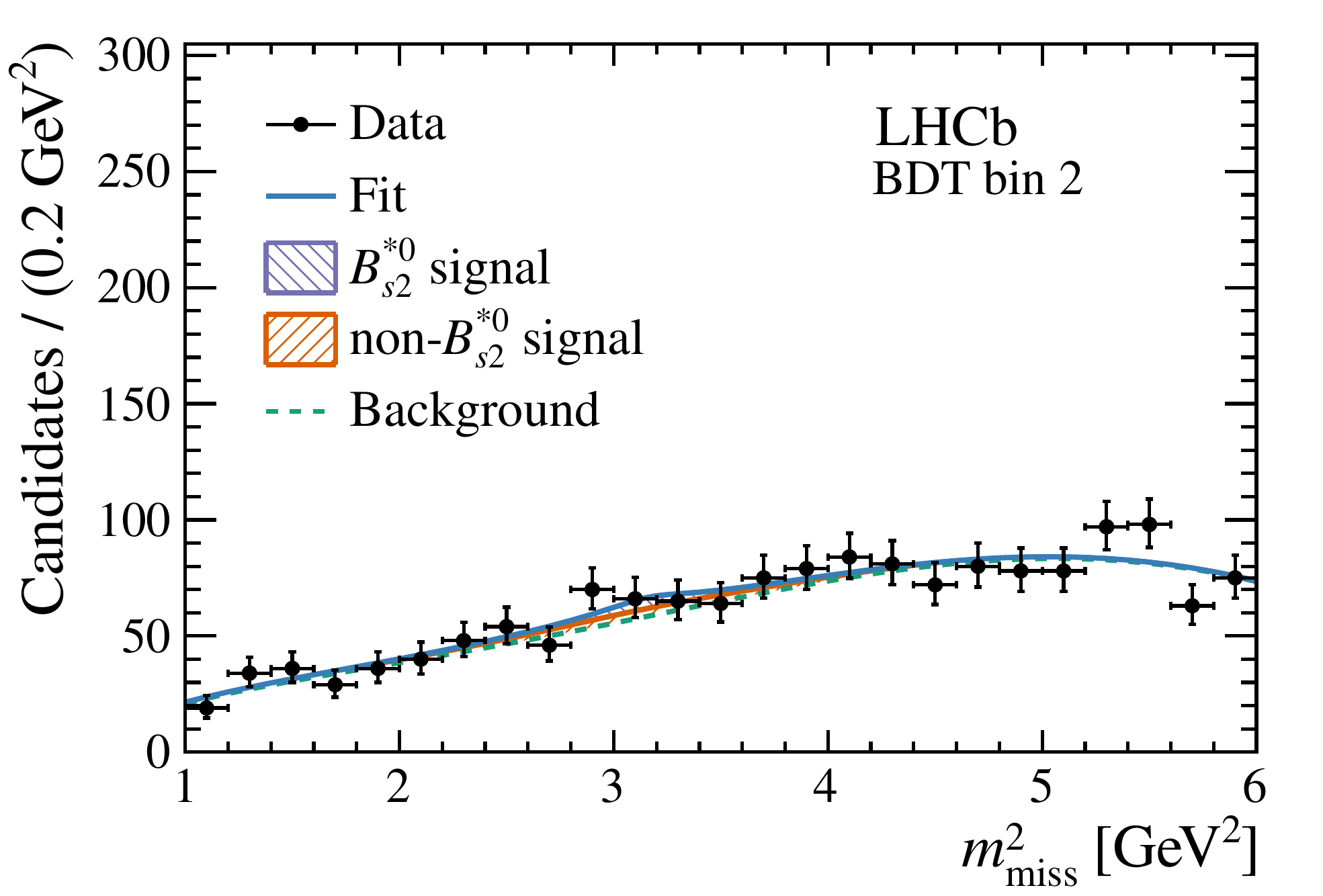}
  \end{sfhalf}
  \begin{sfhalf}
    \includegraphics[width=\textwidth]{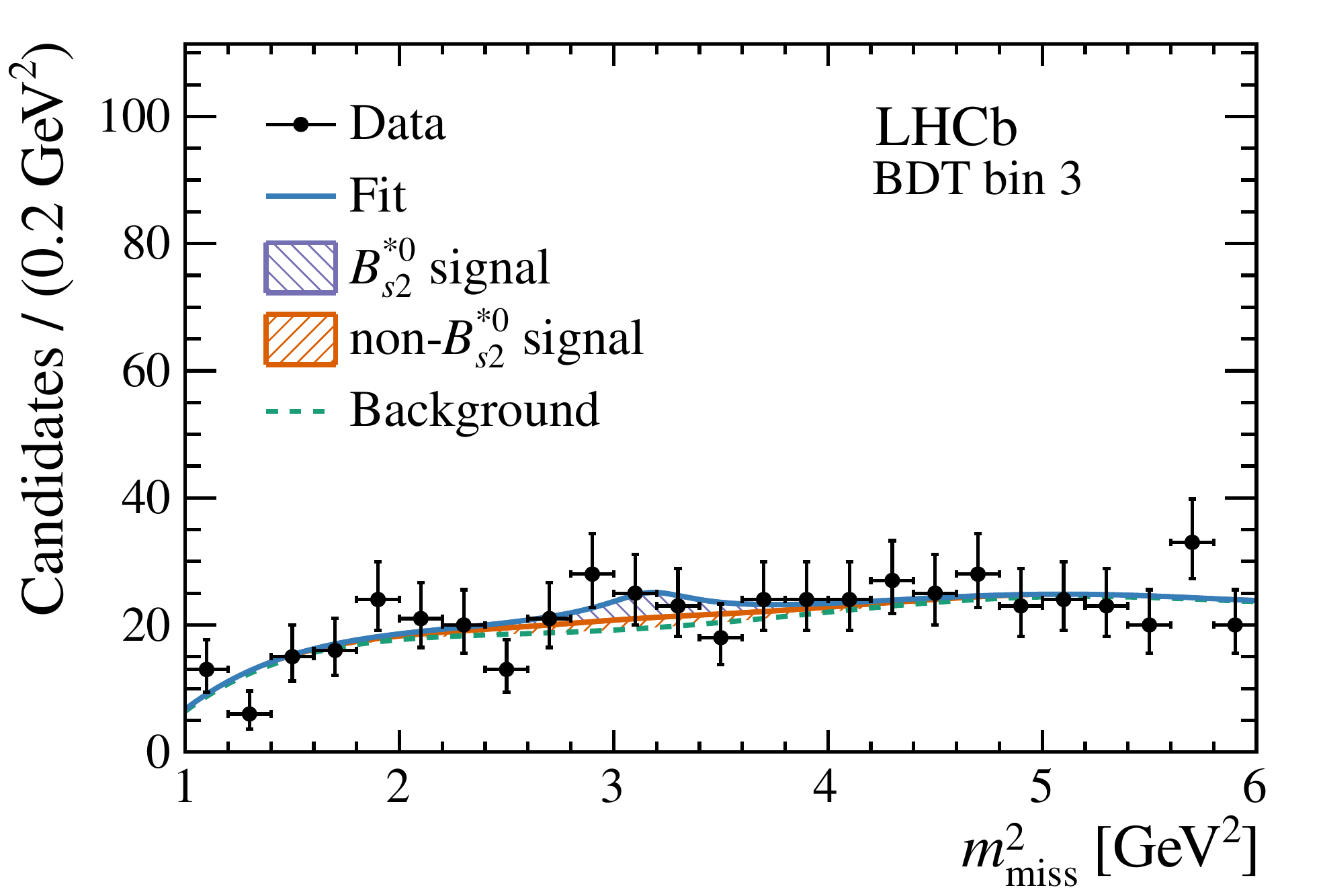}
  \end{sfhalf}
  \begin{sfhalf}
    \includegraphics[width=\textwidth]{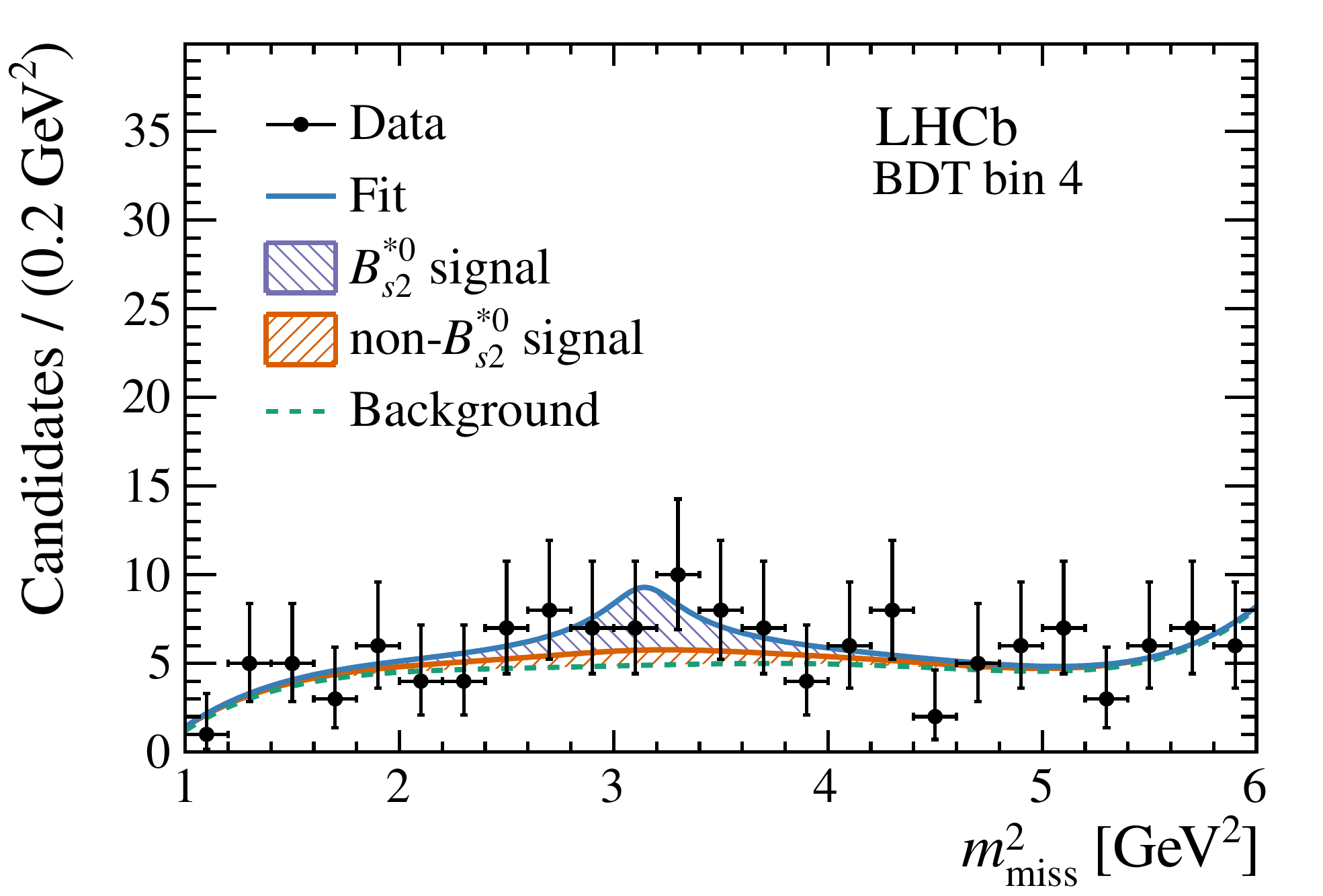}
  \end{sfhalf}
  \caption{Fits to the missing-mass-squared distribution \OSK signal sample in each bin of BDT output included in the final fit. The best fit is overlaid. BDT bin 1 is the most background-like. The fit is performed using a fifth degree polynomial description of the background.\label{fig:fit}}
\end{figure}

\begin{figure}[htbp]
  \begin{center}
    \begin{subfigure}{0.7\textwidth}
      \includegraphics[width=\textwidth]{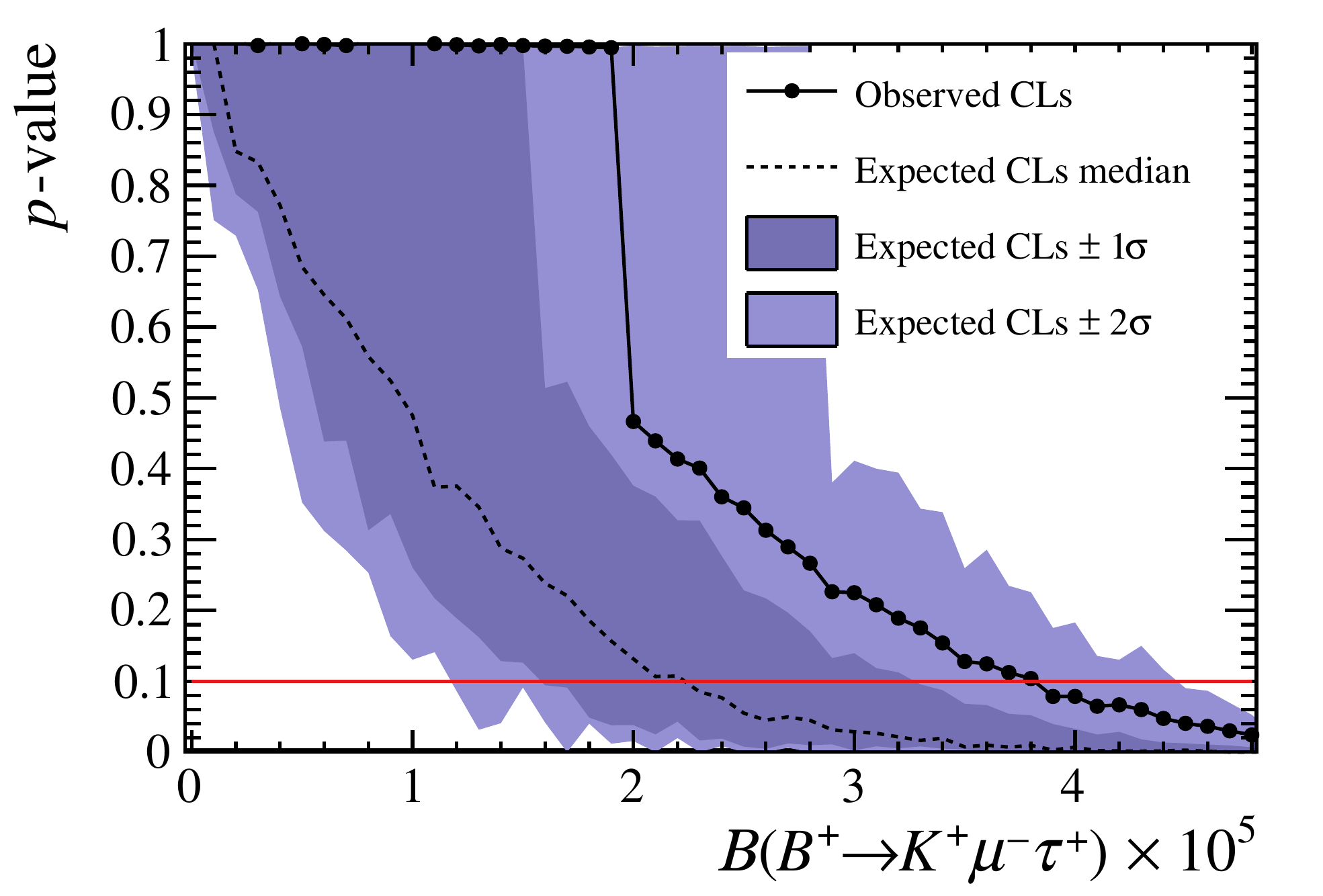}
    \end{subfigure}
  \end{center}
  \caption{Scan of the $p$-value in the signal branching fraction used to determine the CLs upper limits, compared to the expected one. The horizontal red line shows a $p$-value of 0.1, used to define the 90\% CL upper limit.\label{fig:clscan}}
\end{figure}

The result at the best fit point is shown in \cref{fig:fit}. The obtained value for the signal branching fraction from the maximum-likelihood fit is ${\qty(1.9 \pm 1.5 ) \times 10^{-5}}$.  
No significant excess is observed, and we set upper limits on the branching fraction using the CLs method~\cite{CLs}. We perform a scan in the signal branching fraction, obtaining the signal and background $p$-values from the distributions of a one-sided profile-likelihood-ratio test statistic obtained with pseudoexperiments in which we vary the constraints on the systematic uncertainties. The scan used to determine the observed limits, compared to the expected one, is shown in \cref{fig:clscan}. The expected upper limit at 90\% CL is \num{2.3e-5}. The observed 90\% and 95\% CL limits, assuming a phase space signal decay model, are:
\begin{align*}
  \BF\qty(\decay{\Bu}{\Kp\mun\taup}) &< \num{3.9e-5}\text{ at 90\% CL}, \\
  &<\num{4.5e-5}\text{ at 95\% CL}.
\end{align*}
An identical limit is obtained when the decay is generated from the effective operators $\mathcal{O}_9^{(')}$ or $\mathcal{O}_{10}^{(')}$. If instead it is produced from $\mathcal{O}^{(')}_S$ or $\mathcal{O}^{(')}_P$, the obtained limit  is $\BF\qty(\decay{\Bu}{\Kp\mun\taup}) <\num{4.4e-5}$ at 90\% CL and ${} < \num{5.0e-5}$ at 95\% CL.

This is the first result from the LHCb experiment for the lepton-flavour violating decay \decay{\Bu}{\Kp\mun\taup}. By studying \Bu mesons from \bstwost decays, we are able to make the first analysis at LHCb of a \B hadron decay using inclusive $\tau$ decays. This provides complementary information to searches for lepton-flavour violation at LHCb with three-prong $\tau$ decays, for example \decay{B_{(s)}^{0}}{\tau^{\pm}\mu^{\mp}} decays~\cite{LHCb-PAPER-2019-016}. We observe no significant signal, and set an upper limit slightly above that obtained by the BaBar collaboration~\cite{Lees:2012zz}.

%\clearpage
% Comment this in for paper drafts; do not include this in analysis note, conference and figure reports
\section*{Acknowledgements}
%
% These Acknowledgements valid from 3-May-2019
%
\noindent We express our gratitude to our colleagues in the CERN
accelerator departments for the excellent performance of the LHC. We
thank the technical and administrative staff at the LHCb
institutes.
We acknowledge support from CERN and from the national agencies:
CAPES, CNPq, FAPERJ and FINEP (Brazil); 
MOST and NSFC (China); 
CNRS/IN2P3 (France); 
BMBF, DFG and MPG (Germany); 
INFN (Italy); 
NWO (Netherlands); 
MNiSW and NCN (Poland); 
MEN/IFA (Romania); 
MSHE (Russia); 
MinECo (Spain); 
SNSF and SER (Switzerland); 
NASU (Ukraine); 
STFC (United Kingdom); 
DOE NP and NSF (USA).
We acknowledge the computing resources that are provided by CERN, IN2P3
(France), KIT and DESY (Germany), INFN (Italy), SURF (Netherlands),
PIC (Spain), GridPP (United Kingdom), RRCKI and Yandex
LLC (Russia), CSCS (Switzerland), IFIN-HH (Romania), CBPF (Brazil),
PL-GRID (Poland) and OSC (USA).
We are indebted to the communities behind the multiple open-source
software packages on which we depend.
Individual groups or members have received support from
AvH Foundation (Germany);
EPLANET, Marie Sk\l{}odowska-Curie Actions and ERC (European Union);
ANR, Labex P2IO and OCEVU, and R\'{e}gion Auvergne-Rh\^{o}ne-Alpes (France);
Key Research Program of Frontier Sciences of CAS, CAS PIFI, and the Thousand Talents Program (China);
RFBR, RSF and Yandex LLC (Russia);
GVA, XuntaGal and GENCAT (Spain);
the Royal Society
and the Leverhulme Trust (United Kingdom).

\addcontentsline{toc}{section}{References}
%\setboolean{inbibliography}{true}
\bibliographystyle{LHCb}
\bibliography{main,standard,LHCb-PAPER,LHCb-CONF,LHCb-DP,LHCb-TDR,kmutau}

\newpage
% LHCb collaboration author list
% Data extracted on February 28th, 2020 at 4:10pm for reference date 10-Dec-2019
\centerline
{\large\bf LHCb collaboration}
\begin
{flushleft}
\small
R.~Aaij$^{31}$,
C.~Abell{\'a}n~Beteta$^{49}$,
T.~Ackernley$^{59}$,
B.~Adeva$^{45}$,
M.~Adinolfi$^{53}$,
H.~Afsharnia$^{9}$,
C.A.~Aidala$^{80}$,
S.~Aiola$^{25}$,
Z.~Ajaltouni$^{9}$,
S.~Akar$^{66}$
% ...

{\footnotesize \it

% ...
$ ^{9}$Universit{\'e} Clermont Auvergne, CNRS/IN2P3, LPC, Clermont-Ferrand, France\\
% ...
$ ^{25}$INFN Sezione di Milano, Milano, Italy\\
% ...
$ ^{31}$Nikhef National Institute for Subatomic Physics, Amsterdam, Netherlands\\
% ...
$ ^{45}$Instituto Galego de F{\'\i}sica de Altas Enerx{\'\i}as (IGFAE), Universidade de Santiago de Compostela, Santiago de Compostela, Spain\\
% ...
$ ^{49}$Physik-Institut, Universit{\"a}t Z{\"u}rich, Z{\"u}rich, Switzerland\\
% ...
$ ^{53}$H.H. Wills Physics Laboratory, University of Bristol, Bristol, United Kingdom\\
% ...
$ ^{59}$Oliver Lodge Laboratory, University of Liverpool, Liverpool, United Kingdom\\
% ...
$ ^{66}$Los Alamos National Laboratory (LANL), Los Alamos, United States\\
$ ^{67}$Syracuse University, Syracuse, NY, United States\\
% ...
$ ^{80}$University of Michigan, Ann Arbor, United States, associated to $^{67}$\\
\bigskip
% ...
\medskip
}
\end{flushleft}

\end{document}